\documentclass[
reprint,
amsmath,
amssymb,
aps,
prx,
longbibliography,
showpacs,
floatfix
]{revtex4-2}
\usepackage{hyperref}
\usepackage{graphicx}
\usepackage{xcolor} 
\usepackage{booktabs}
\usepackage{mathtools}
\usepackage{tabularx}
\usepackage{multirow}

\usepackage{braket}

\usepackage{comment}
\usepackage{enumitem}
\usepackage{soul}

\definecolor{color1}{rgb}{0,0.25,0.70}
\hypersetup{colorlinks=true,
    linkcolor={color1},
    citecolor={color1},
    urlcolor={color1}
}
\def\Circlearrowleft{\ensuremath{%
\scalebox{1.2}{\textcolor{black}{\rotatebox[origin=c]{250}{$\circlearrowleft$}}}}}

\begin{document}

\title{Shift and Polarization of Excitons from Quantum Geometry}

\author{Carolina Paiva}
\email{paiva@mail.tau.ac.il}

\author{Tobias Holder}

 \author{Roni Ilan}
 \affiliation{School of Physics and Astronomy, Tel Aviv University, Tel Aviv 6997801, Israel}

\date{\today}

\begin{abstract}
Despite a long history, certain aspects of excitons - the bound inter-band states which form when a valence band hole and a conduction band electron pair - have remained relatively unexplored. 
This holds particularly true for the wavefunction of an exciton, for which few properties have been explored theoretically in various limiting cases.
An intuitive language robustly characterizing the topology of bound electron-hole states is lacking, but needed in order to address the global features of the charge distribution of the excitonic state, to properly understand their transport theory, and to supplement the numerical investigation of excitons in ab-initio approaches.
Here, we address these gaps by developing a comprehensive framework for the quantum geometry and topology of two-dimensional exciton states in terms of the exact connections which describe the interaction-renormalized exciton bundle in a periodic lattice. 
Based on this description, we derive two gauge-invariant quantities, which we identify as the exciton shift vector and the exciton dipole vector. 
Using the shift vector, we elucidate the topology of exciton bands compared to the topology of the parent electronic band structure, pinpointing precisely how interactions can introduce nontrivial topology to the exciton bands beyond the topology which is contained in the single-particle bands.
We further elucidate how shift and polarizations enter into the semiclassical equations of motion for the exciton center of mass coordinates.
\end{abstract}

\maketitle

\section{Introduction}
Recent progress in the fabrication and control of layered materials in the mono- or few-layer limit have reinvigorated the study of the fundamental properties of excitonic states~\cite{Wilson2021,Regan2022,Ciarrocchi2022}. This is not only due to the increased exciton stability and lifetime in these quasi-two dimensional systems, but also thanks to the added tunability of the electronic properties by strain or gating~\cite{Mueller_2018,Rivera2015,Rivera2016,Rickhaus_2021}. For example, in transition metal dichalcogenides (TMDs) it has become possible to systematically explore mechanisms of exciton formation~\cite{Diana_2015,Trovatello2020,Jiang2021} and dynamics~\cite{Jin_2018,Esteve_2024,Jiang2021}. Moreover, moiré materials like twisted bilayer graphene (TBG) or twisted TMD layered structures, have likewise been demonstrated to exhibit novel exciton bands~\cite{MacDonald_2017,Patel_2019,Andersen_2021,Jin_2018}.
However, while the binding energy and optical activity of excitons has been studied in excruciating detail, much less is known about the topological and quantum geometric properties of the excitonic band structure.
Recent works have started to characterize exciton topology, suggesting that an appropriately generalized Berry curvature can be defined for excitons, which will for example affect the exciton spectrum~\cite{Franz_2011,Zhou_2015,Srivastava_2015}. Additionally, optical selection rules which determine whether excitonic processes are active (bright excitons) or inactive (dark excitons), are altered by this exciton Berry curvature~\cite{Louie_2018,Zhang_2018,Xu_2020}. 
Finally, it has been suggested that topology can affect exciton dynamics and transport, notably reporting an exciton Hall effect under the influence of a magnetic field~\cite{Niu_2008,Fertig_2021,Chaudhary_2021,Kwan_2021,Arnardottir_2012,Kozin_2021,Xie2023,Beaulieu2024}. 
Some works have also discussed the role of quantum metric~\cite{Hu2022} not only in exciton transport in flat band systems~\cite{ying2024}, but also in bounds on the exciton wavefunction spread and phase stiffness of interlayer exciton condensate~\cite{Verma_2024}.

Despite these advances, a comprehensive framework of the quantum geometry and topology of exciton states is currently lacking. In particular, it is unclear how the topology of the underlying electron and hole bands are related to the resulting topology of the exciton band structure in the presence of interaction. This situation is exacerbated by the dichotomy of microsopic detail that is typically employed: Numerical  studies based on ab-initio methods provide an overwhelming amount of information, making it challenging to disentangle specific contributions. On the other hand, low-energy effective models almost universally  use the effective mass approximation to obtain analytical solutions, and thus by construction cannot account for features determined by global topology.

In this work, we remedy this disconnect by deriving and characterizing the geometric properties of excitonic states based on the full lattice wavefunctions. In particular, we clarify how the electron-hole interaction can affect the topology of the bound state. This becomes possible in a generalized framework which entirely sidesteps the effective mass approximation. As our main result, we identify two gauge-invariant quantities that fully characterize the exciton's topology and geometry: The \emph{exciton shift vector} $\mathbf{S}_{\mathbf{Q}}$ associated with the center-of-mass momentum $\mathbf{Q}$ and the \emph{exciton dipole vector} $\mathbf{P}_{\mathbf{p}}$, which is connected to the electron-hole relative momentum $\mathbf{p}$. The shift vector encodes how the electron-hole interaction influences the exciton band topology, which makes it possible to fully characterize the latter. 
On the other hand, the dipole vector is a polarization vector associated with the exciton's intrinsic dipole moment, which couples to an external electric field. $\mathbf{P}_{\mathbf{p}}$ thus plays a crucial role in exciton dynamics, giving rise to a new anomalous velocity term in the exciton's semiclassical equations of motion. 

This paper is organized as follows: In Sec.~\ref{sec:Exciton_Wavefunction_beyond_the_Effective_Mass_Approximation}, we introduce a formalism that sidesteps the effective mass approximation, enabling the study of exciton band geometry and topology across the entire Brillouin zone. Sec.~\ref{sec:singular_connection} establishes the mathematical framework for the concept of singular connection, distinct from the Berry connection but useful for discussing pairs of quantum states~\cite{Mera_2021}. In Sec.~\ref{sec:Shift_Vector}, we define the exciton shift vector, a gauge-invariant quantity encapsulating exciton topology. Sec.~\ref{sec:relative_polarization} introduces the polarization of the exciton, related to its relative momenta. Finally, in Sec.~\ref{sec:semiclassical_equations_motion}, we explore the impact of this polarization on exciton dynamics before concluding in Sec.~\ref{sec:conclusion}.

\section{Exciton Wavefunction Beyond the Effective Mass Approximation}
\label{sec:Exciton_Wavefunction_beyond_the_Effective_Mass_Approximation}

In a crystalline system where the Bloch theorem applies, the single particle electron and hole states, which belong to the valence and conduction band, respectively, take the Bloch form $\psi^{n}(\mathbf{k})=e^{i\mathbf{k}\cdot\mathbf{r}}u^{n}_{\mathbf{k}}(\mathbf{r})$, where $u^{n}_{\mathbf{k}}(\mathbf{r})$ is a function that obeys the periodicity of the lattice, and $n$ is the band index. In our case, the band index will be labelled as $v$ or $c$ for valence or conduction, respectively, and $\mathbf{k}$ is the single particle lattice momentum. The exciton wavefunction is a two-particle state that can be written as a superposition of products of the particle and the hole states
\begin{equation}
\psi^{exc} = \sum_{\mathbf{k}^c,\mathbf{k}^v} F(\mathbf{k}^c,\mathbf{k}^v) e^{i\mathbf{k}^c\cdot\mathbf{r}_c}e^{-i\mathbf{k}^v\cdot\mathbf{r}_v}u^{c}_{\mathbf{k}^c}(\mathbf{r}_c)(u^{v}_{\mathbf{k}^v}(\mathbf{r}_v))^{*},
\end{equation}
where $\mathbf{k}^{c}$, $\mathbf{k}^{v}$ are the momentum of the electron and the hole, respectively, and run over the entire Brillouin zone. The coefficients $F(\mathbf{k}^c,\mathbf{k}^v)$ determine the structure of the exciton wavefunction. 

Translation symmetry dictates that the motion of the exciton can be decomposed into the relative motion of an electron and a hole, and the center of mass motion. So, the more natural set of coordinates to work with are the center of mass momentum $\mathbf{Q}$ and relative momentum $\mathbf{p}$. In the effective mass approximation, this decomposition is  known from the treatment of two particle systems: the center of mass position $\mathbf{R}=\gamma_{e}\mathbf{r}_{e}+\gamma_{h}\mathbf{r}_{h}$ is defined with the relative masses of the electron and hole, respectively given by $\gamma_{e}=m_e/(m_e+m_h)$ and $\gamma_{h}=m_h/(m_e+m_h)$, with $m_e$ and $m_h$ the electron and hole mass, respectively. The center of mass momentum $\mathbf{Q}$ is naturally defined in terms of $\mathbf{R}$ as $\mathbf{Q}=(m_e+m_h)\Dot{\mathbf{R}}$. The relative coordinate $\mathbf{r}=\mathbf{r}_e-\mathbf{r}_h$ is simply the difference between the position of the electron and hole. Then, the relative momentum $\mathbf{p}=\mu \Dot{\mathbf{r}}$ will be the time derivative of the relative motion coordinate, with $\mu=m_em_h/(m_e+m_h)$ the reduced mass of the system.
However, as explained before, the treatment of the exciton in the effective mass approximation is restrictive and confines any prediction to be relevant only for confined regions in reciprocal space. One of the main points of our work is that this approximation is in fact unnecessary. Indeed, one can perform a generic coordinate transformation from the coordinates of the electron and hole, $(\mathbf{r}_c,\mathbf{r}_v,\mathbf{k}^{c},\mathbf{k}^{v})$, to some other set of coordinates of the composite system, $(\mathbf{r},\mathbf{R},\mathbf{p},\mathbf{Q})$. As we explicitly show in the appendix, any choice for this coordinate transformation is valid (as long as it is invertible), because the Berry curvature (the 2-form object) as well as the observables of the system are independent of the coordinate basis in which we choose to work. 

Since the choice of the coefficients in the change of basis from $(\mathbf{r}_c,\mathbf{r}_v,\mathbf{k}^{c},\mathbf{k}^{v})$ to $(\mathbf{r},\mathbf{R},\mathbf{Q}, \mathbf{p})$ will not change the Berry curvature or the observables of the system that are computed from the Berry curvature, in the following we  motivate a simple and symmetric form of this coordinate transform. The names of the variables will be kept the same as the ones used in the effective mass approximation, i.e. $\mathbf{R}$ is the ``center of mass position", $\mathbf{r}$ is the relative coordinate, $\mathbf{p}$ is the relative momentum and $\mathbf{Q}$ is the ``center of mass momentum".
When choosing the coefficients, we wish to ensure that in the exciton wavefunction, the center of mass momentum, $\mathbf{Q}$, appears only in inner products with the center of mass position $\mathbf{R}=(\mathbf{r}_c+\mathbf{r}_v)/2$. Analogously, the relative momenta $\mathbf{p}$ will only appear in inner products with the relative coordinate $\mathbf{r}=\mathbf{r}_c-\mathbf{r}_v$. We therefore choose to define the center of mass momentum as $\mathbf{Q}=\mathbf{k}^c-\mathbf{k}^v$ and the relative momentum as $\mathbf{p}=\frac{\mathbf{k}^c+\mathbf{k}^v}{2}$. Note that this notation is consistent with the fact that in the literature, a vertical excitation of an electron, which leaves a hole behind, is associated with an exciton with zero center of mass momentum, $\mathbf{Q}=0$.

We note in passing that the Coulomb interaction, responsible for the binding of the exciton, conserves the total momentum of the system. Therefore, the two summations in $\mathbf{k}^c$ and $\mathbf{k}^v$ can be reduced to a single sum over $\mathbf{p}$, and a sum over a delta function in $\mathbf{Q}-\mathbf{k}^c+\mathbf{k}^v$. By evaluating the latter, the coordinate transformation $(\mathbf{k}^{c},\mathbf{k}^{v})$ to $(\mathbf{Q}, \mathbf{p})$ is no longer manifestly invertible. The wavefunction of the exciton can then parameterized by the center of mass momentum $\mathbf{Q}$ 
\begin{align}
\psi^{exc}_{\mathbf{Q}} &= e^{i\mathbf{Q}\cdot \mathbf{R}}\sum_{\mathbf{p}} F(\mathbf{p},\mathbf{Q})e^{i\mathbf{p}\cdot \mathbf{r}}u^{c}(\mathbf{p}+\mathbf{Q}/2)(u^{v}(\mathbf{p}-\mathbf{Q}/2))^{*} \nonumber\\
\label{eq:exciton_wavefunction}
&= e^{i\mathbf{Q}\cdot \mathbf{R}} U^{exc}(\mathbf{Q}),
\end{align}
where trivially $\psi^{exc}=\sum_{\mathbf{Q}}\psi^{exc}_{\mathbf{Q}}\delta(\mathbf{Q}-\mathbf{k}^c+\mathbf{k}^v)$.
In writing Eq.~\ref{eq:exciton_wavefunction}, $\mathbf{Q}$ becomes a quantum number for the dynamics of the exciton as a quasi-particle, while $\mathbf{p}$ is an internal degree of freedom of the exciton, to be summed over.
Before moving on to the geometrical aspects of the exciton wavefunction, we note that the coefficients $F(\mathbf{p},\mathbf{Q})$ are obtained by solving the Bethe Salpeter equation~\cite{Louie_2000}, given by
\begin{align}
\epsilon^{exc}_{\mathbf{Q}} F\mathbf{(p,Q)}
&=(E^c_{\mathbf{p}+\mathbf{Q}/2}-E^v_{\mathbf{p}-\mathbf{Q}/2})F\mathbf{(p,Q)}
\nonumber\\
&\quad+\int d^dp'K^{eh}_{\mathbf{p},\mathbf{p'}}F\mathbf{(p',Q)},
\label{eq:BSE}
\end{align}
where the integration runs over the entire Brillouin zone. Here $E^{c},E^{v}$ are the conduction and valence band energies at the corresponding momentum, and $\epsilon^{exc}_{\mathbf{Q}}$ is the exciton energy. The interaction kernel $K^{eh}_{\mathbf{p},\mathbf{p'}}$ is the matrix element of the Coulomb interaction, which contains a direct electron-hole attractive screened term and an exchange repulsive term.

\section{Singular Connection vs. Berry connection}
\label{sec:singular_connection}
 
In this section, we aim to determine how the interaction between electrons and holes influences the topology of exciton bands. The topology of two independent particles has been extensively studied in the literature~\cite{Mera_2021}. However, it is far from clear how one should treat the case of a bound state formed from two originally independent particles interacting through a Coulomb interaction. Our  key idea is to find a geometric quantity that captures the role of the interaction and decouples the interaction effect from the single band contributions to the topology of the exciton. It turns out that there is a useful but not so well-known geometric quantity called the singular connection that does exactly that. We show that the singular connection depends solely on the coefficients $F(\mathbf{p},\mathbf{Q})$, which have a nontrivial momentum dependence only when interactions are turned on, and hence captures this effect. This connection allows us to construct the shift vector of the exciton, which contains all the information about its topological properties, allowing for a distinction between trivial and topological excitons.

We first provide a concise yet comprehensive mathematical introduction to singular connections, contrasting them with the more well known Berry connection. The fluent reader familiar with the concept of singular connections may skip this section. In simple terms, the singular connection is an alternative connection to the usual Berry connection for a single or a pair of quantum states, and even though it behaves in a singular manner over the Brillouin zone, it still defines a valid connection capturing the topology of the system. As an example, it was shown in~\cite{Mera_2021} that this connection appears naturally in the context of resonant optical responses describing optical transitions between a pair of quantum states belonging to different energy bands. 

The geometry of band theory hinges on mathematical constructs known as vector bundles~\cite{Nakahara_2003,Fruchart_2013,Kobayashi_2014,Rui_loja_24}. Essentially, a vector bundle $\mathcal{E}$ consists of a base manifold $B$ with a fiber $E$ attached to each point of $B$. These fibers can be real or complex vector spaces. Formally, the vector bundle $\mathcal{E}$ is a collection of fibers isomorphic, but not in a canonical way, to a typical fiber E (represented in Eq.\eqref{eq:schematic_bundle} by the arrow $\xhookrightarrow{}$), with a projection map $\pi$ from $\mathcal{E}$ to the base space $B$ (represented in Eq.\eqref{eq:schematic_bundle} by the arrow $\xlongrightarrow[]{\pi}$). Additionally, there can also be a group $G$ that acts on the fibers $E$ (represented in Eq.\eqref{eq:schematic_bundle} by the arrow $\Circlearrowleft$):
\begin{equation}
 G \Circlearrowleft E \xhookrightarrow{} \mathcal{E} \xlongrightarrow[]{\pi} B.
 \label{eq:schematic_bundle}
\end{equation}
The geometry of a vector bundle is described by objects like connections and curvature, while its topology is characterized by topological invariants, the characteristic classes, with the Chern number being a well-known example. In the context of band theory of a single electron in a periodic potential the relevant vector bundles are called Bloch bundles. Here, the Brillouin zone $BZ^2$, which is topologically equivalent to a torus, serves as the base space while the fibers are complex vector spaces, which represent the Hilbert spaces of the energy bands of the system. The dimension of these depends on the total number of bands $n$. Although the complete Hilbert space forms a vector bundle, it's always a trivial one, and so the total Berry curvature sums to zero when all the energy bands are taken into account. This triviality is a familiar concept in condensed matter physics.

A more intriguing scenario arises when considering subsets of bands, such as a one-dimensional Hilbert space corresponding to a single band. Quantum states of this band, are denoted by $\ket{\psi(\mathbf{k})}\in \mathbb{C}^{n}$, for the rest of the discussion we assume them to be normalized. The geometry of these quantum states, which are formally defined up to multiplication by a non-vanishing scalar, of a single isolated band is characterized by a smooth map $f:BZ^2\rightarrow \mathbb{C}P^{n-1}$, where $\mathbb{C}P^{n-1}$ is the ($n-1$)-dimensional complex projective space consisting of one-dimensional linear subspaces of $\mathbb{C}^{n}$. To clarify, the map $f$ locally assigns to each point in the Brillouin zone $\mathbf{k}$, a vector $\ket{\psi(\mathbf{k})}$ up to a scale. Using this map we are able to define a line bundle $\mathcal{L}\rightarrow BZ^2$, whose fiber at each $\mathbf{k}$ is the 1-dimensional vector subspace of $\mathbb{C}^{n}$ determined by $f(\mathbf{k})$. 

A crucial concept in the context of line bundles is that of a section. A section of a line bundle $\mathcal{L}$ is a map $s:\mathbf{k}\mapsto s(\mathbf{k})$, where $s(\mathbf{k})$ is in the fiber of $\mathcal{L}$ over $\mathbf{k}$. Typically, sections are only defined on an open subset in the Brillouin zone $BZ^2$ these are called local sections. A particular choice of local representation $\ket{\psi(\mathbf{k})}$ of the map $f$, valid in some open subset of the $BZ^2$, gives a local section of the line bundle $\mathcal{L}$. However, there can also be sections that are defined over the entire Brillouin zone $BZ^2$, these will be global sections. The latter play a crucial role in determining whether a vector bundle is trivial. Specifically, the existence of a global and nonvanishing section over $BZ^2$ signifies the triviality of the vector bundle. Conceptually, a section can be understood as a generalization of a function defined over the Brillouin zone. In the case of a trivial vector bundle, a global section behaves like an ordinary function. Conversely, in a twisted bundle, the section is a function that is defined over different patches that together cover the Brillouin zone, appearing as a multi-valued (and hence ill-defined) function. An interesting scenario arises with twisted vector bundles, where an obstruction exists in finding a global section that is non-vanishing. However, as it turns out, even in such cases it is possible to find a global section by allowing it to vanish in certain parts of the Brillouin zone $BZ^2$. These global sections will prove pivotal in defining singular connections.

Numerous connections can be constructed on a line bundle. A connection is characterized by local one-forms defined on open subsets of the Brillouin zone $BZ^2$. These connection one-forms must satisfy the following property: as we change the local representation of the map $f$, as $\ket{\psi(\mathbf{k})}\rightarrow g(\mathbf{k})\ket{\psi(\mathbf{k})}$, with $g(\mathbf{k})$ representing a local gauge transformation, the one-forms transform according to:
\begin{equation}
    \mathcal{A}(\mathbf{k})\rightarrow \mathcal{A}(\mathbf{k})-i g(\mathbf{k})^{-1}\mathrm{d}g(\mathbf{k}).
    \label{eq:gauge_tranformation_connection}
\end{equation}
The well-known Berry connection,  usually denoted in the physics literature as $\mathcal{A}_{Berry}=-i \bra{\psi(\mathbf{k})}\mathrm{d}\ket{\psi(\mathbf{k})}$, and first introduced by Michael Berry in~\cite{Berry_1984}, is one of the connections that obeys the condition above. Remarkably, this connection emerges not only in condensed matter systems but also in various other quantum mechanics contexts. The reason for this is that its construction relies solely on the information provided by the map $f$ and  the line bundle constructed from it $\mathcal{L}$. From this connection it is possible to construct the 2-form called the Berry curvature, $\Omega_{Berry}=\mathrm{d}\mathcal{A}_{Berry}$. The curvature of a connection on the line bundle $\mathcal{L}$, in particular the Berry curvature, determines completely its first Chern class, which is a characteristic class. Integration of the first Chern class over the entire Brillouin zone $BZ^2$ yields an integer known as the Chern number, a fundamental topological invariant.

Alternative connections apart from the Berry connection could have been chosen in order to characterize band geometry and topology. Suppose we have two connections $\nabla$ and $\widetilde{\nabla}$ locally described by gauge fields $\mathcal{A}$ and $\widetilde{\mathcal{A}}$, respectively, both transforming under gauge transformation as depicted in Eq.~\eqref{eq:gauge_tranformation_connection}. Taking the difference $S=\nabla-\widetilde{\nabla}$, which is locally determined by $\mathcal{A}-\widetilde{\mathcal{A}}$, yields a gauge invariant, and hence, globally defined 1-form. Consequently, the difference in the curvatures $\mathrm{d}S=\Omega-\widetilde{\Omega}$ is an exact 2-form, implying its integral over the entire Brillouin zone $BZ^2$ vanishes. This observation allows us to conclude that the first Chern class remains independent of the chosen connection. Nonetheless, singular connections, as explained in~\cite{Mera_2021}, might contain interesting information about physical systems. Before delving into their construction for excitons, we summarize the main aspects in the single-particle picture. 

The singular connection stands apart from the Berry connection in its requirement for a global section $s$ of the line bundle. A generic global section is expressed in terms of the local representation of the fiber $\ket{\psi(\mathbf{k})}$ at $\mathbf{k}$ as $s(\mathbf{k})=t(\mathbf{k})\ket{\psi(\mathbf{k})}$, where $t(\mathbf{k})$ is a smooth complex function over the domain where $\ket{\psi(\mathbf{k})}$ is defined. Combining these local representations forms the global section. It's crucial for global sections to be gauge invariant; under a gauge transformation of the local representation $\ket{\psi(\mathbf{k})}\rightarrow g(\mathbf{k})\ket{\psi(\mathbf{k})}$ the coefficient transforms in the opposite way as $t(\mathbf{k})\rightarrow g(\mathbf{k})^{-1}t(\mathbf{k})$. The singular connection is defined  based the global section $s$ as

\begin{equation}
    \nabla s = 0.
    \label{eq:singular_connection_def}
\end{equation}
Using the local representation $s(\mathbf{k})=t(\mathbf{k})\ket{\psi(\mathbf{k})}$ and after some algebra we find that
\begin{equation}
    \nabla \ket{\psi(\mathbf{k})} = i\mathrm{d} \log{(t(\mathbf{k}))} \ket{\psi(\mathbf{k})}.
    \label{eq:connection_singular}
\end{equation}
In the gauge specified by $\ket{\psi(\mathbf{k})}$, the connection 1-form is determined by how the connection acts on $\ket{\psi(\mathbf{k})}$. From Eq.~\eqref{eq:connection_singular} one obtains that a connection 1-form of the singular connection is given by
\begin{equation}
    \mathcal{A}_{sing}(\mathbf{k}) = i\mathrm{d}\log{(t(\mathbf{k}))}.
    \label{eq:singular_form}
\end{equation}
The term ``singular" aptly describes the connection's behavior; it becomes singular whenever $t(\mathbf{k})=0$, indicating the vanishing of the global section $s$. It is noteworthy that for a given line bundle, multiple global sections exist, and the choice of section impacts the resulting singular connection. However, as previously explained, all choices remain valid for studying the system's topology. Yet, there may be physical motivations behind selecting a particular section, rendering the resulting singular connection physically relevant. The curvature of a singular connection is also singular and is referred to as singular curvature. One can prove~\cite{Mera_2021} that by writing $t(\mathbf{k})=x_{1}(\mathbf{k})+ix_{2}(\mathbf{k})$ the singular curvature takes the following form:
\begin{equation}
    \Omega_{sing}(\mathbf{k}) = \mathrm{d}\mathcal{A}_{sing}(\mathbf{k}) = -2\pi \delta^{2}(t(\mathbf{k}))\mathrm{d}x_{1}(\mathbf{k})\wedge \mathrm{d}x_{2}(\mathbf{k}).
    \label{eq:singular_curvature_general}
\end{equation}
Therefore, the singular curvature localizes (takes an infinite value in the sense of the Dirac delta function) over the set $N_V=\{\mathbf{k}\in BZ^2 : t(\mathbf{k})=0\}$, in which the global section used for its construction becomes zero.
\section{Singular connection of excitons}
Given this background, we are now in the position to discuss the topology of excitons. The exciton is a quasi-particle that is formed by an electron and hole that interact through Coulomb interaction forming a bound state. As it was explained before, our main goal is to quantify the impact of electron-hole interactions encapsulated within the coefficients $F(\mathbf{p},\mathbf{Q})$. To achieve this, we'll initially explore the geometries of electrons and holes as independent entities.

The general discussion of the line bundle for two independent particles can for example be found in~\cite{Mera_2021}. In order to construct line bundles specific to electrons and holes, we begin by considering two distinct families of quantum states: $\ket{u^{c}(\mathbf{p}+\mathbf{Q}/2)}$ and $\ket{u^{v}(\mathbf{p}-\mathbf{Q}/2)}$. In discussing exciton topology, we treat the exciton as a quasi-particle with momentum dependence in $\mathbf{Q}$. For this reason and for the rest of the section, we will fix $\mathbf{p}$ take $\mathbf{p}=0$. The internal momentum $\mathbf{p}$ will be reintroduced in Sec.~\ref{sec:relative_polarization}. 
The state $\ket{u^{c}(\mathbf{Q}/2)}$ corresponds to electron states residing in the conduction band, while $\ket{(u^{v}(-\mathbf{Q}/2))^{*}}$ represents hole states. Consequently, we establish corresponding line bundles for the conduction and valence bands, denoted as $\mathcal{L}_{c}\rightarrow BZ^2$ and $\mathcal{L}_{v}\rightarrow BZ^2$, respectively. 
Here, $BZ^2$ denotes the Brillouin zone of the exciton, which is four times the size of the original Brillouin zone. In the simplest case, this means that the lattice momenta $\mathbf{Q}$ range from $-2\pi$ to $2\pi$. 
The fibers of $\mathcal{L}_{c}$ and $\mathcal{L}_{v}$ consist of one dimensional complex vector spaces spanned by $\ket{u^{c}(\mathbf{Q}/2)}$ and $\ket{u^{v}(-\mathbf{Q}/2)}$, respectively.

From these individual line bundles, we construct the homomorphism bundle  $\mathcal{L}_{0}:=\mathcal{L}_{c}\otimes \mathcal{L}^{*}_{v}=$Hom($\mathcal{L}_v,\mathcal{L}_c$), with $\mathcal{L}^{*}_{v}$ the dual bundle of $\mathcal{L}_{v}$, whose fibers at each $\mathbf{Q}\in BZ^2$ is the dual of the fiber of $\mathcal{L}_{v}$ at the same quasi-momenta. The fibers of $\mathcal{L}_{0}$ consist of one dimensional complex vector space spanned by $\ket{u^{c}(\mathbf{Q}/2)}\ket{(u^{v}(-\mathbf{Q}/2))^{*}}$. Moreover, $\mathcal{L}_{0}$ inherits a connection induced from the Berry connections on $\mathcal{L}_c$ and $\mathcal{L}_v$. In the local gauge specified by $\ket{u^{c}(\mathbf{Q}/2)}\ket{(u^{v}(-\mathbf{Q}/2))^{*}}$, the connection 1-form is determined by:
\begin{equation}
    \mathcal{A}^{cv}:=\mathcal{A}^{c}-\mathcal{A}^{v}:=\mathcal{A}^{c}_{Berry}-\mathcal{A}^{v}_{Berry},
    \label{eq:Berry_Connection_L0}
\end{equation}
with $\mathcal{A}^{c}_{Berry}=-i\bra{u^{c}(\mathbf{Q}/2)}\mathrm{d}\ket{u^{c}(\mathbf{Q}/2)}$ and $\mathcal{A}^{v}_{Berry}=-i\bra{u^{v}(-\mathbf{Q}/2)}\mathrm{d}\ket{u^{v}(-\mathbf{Q}/2)}$ denoting the local connection 1-forms for the Berry connections of the conduction and valence bands, respectively.

The construction of the exciton bundle follows a similar procedure. Before explaining the construction, we remind the reader that we fixed the value of $\mathbf{p}$ at the beginning of the section. 
Notice that a different choice of $\mathbf{p}$ at the single particle level can be viewed as a translation in the center of mass momentum, $\mathbf{Q}$ by $2\mathbf{p}$ for the conduction band and by $-2\mathbf{p}$ for the valence band.
Since translations in momentum space do not alter the topological invariants of the system, any choice of $\mathbf{p}$ is equally valid. So, $\mathbf{p}=0$ is chosen for convenience.

Next, we define a family of quantum states $\ket{U^{exc}(\mathbf{Q})}$,
 representing excitonic states whose wavefunctions we introduced in the previous section in Eq.~\eqref{eq:exciton_wavefunction}. Subsequently, we construct the corresponding line bundle denoted as $\mathcal{L}_{int}\rightarrow BZ^2$, with $BZ^2$ the Brillouin zone of the exciton. We equip this line bundle with the Berry connection. In the local gauge specified by $\ket{U^{exc}(\mathbf{Q})}$, the connection 1-form is determined by:
\begin{equation}
\mathcal{A}^{exc}:=\mathcal{A}^{exc}_{Berry} = -i\bra{U^{exc}(\mathbf{Q})}\mathrm{d}\ket{U^{exc}(\mathbf{Q})}.
\label{eq:Berry_Connection_Exciton}
\end{equation}

Before proceeding, it is important to address concerns about the implicit summation over $\mathbf{p}$ in Eq.~\eqref{eq:exciton_wavefunction} and, by consequence, in Eq.~\eqref{eq:Berry_Connection_Exciton}. The key point is that $\ket{U^{exc}(\mathbf{Q})}$ serves as a section of a line bundle, $\mathcal{L}_{int}$, which transforms independently of $\mathbf{p}$ under gauge transformations.  The summation over the internal variable $\mathbf{p}$ does not affect the topology of the exciton, as it depends solely on $\mathbf{Q}$.

We now consider the line bundles $\mathcal{L}_{0}$ and $\mathcal{L}_{int}$. From these, we define the homomorphism bundle $\mathcal{L}_{F}:=\mathcal{L}_{0}\otimes (\mathcal{L}_{int})^{*}=$Hom($\mathcal{L}_{int},\mathcal{L}_{0}$), one important point to consider is that the geometric quantites of the line bundle $\mathcal{L}_{F}$ will depend on the specific choice of $\mathbf{p}$, however the topology will not be affected by this choice.  This bundle's fiber at $\mathbf{Q}\in BZ^2$ comprises the one-dimensional complex vector space spanned by $\ket{u^{c}(\mathbf{Q}/2)}\ket{(u^{v}(-\mathbf{Q}/2))^{*}}\ket{(U^{exc}(\mathbf{Q}))^{*}}$. By construction, $F(\mathbf{p}=0,\mathbf{Q}):=F(\mathbf{Q})$ defines a global section of $\mathcal{L}_{F}$. Indeed, by explicitly checking how the section transforms upon the gauge transformations: $\ket{u^{c}_{\mathbf{k}}}\rightarrow e^{i\varphi^{c}(\mathbf{k})}\ket{u^{c}_{\mathbf{k}}}$, $\ket{u^{v}_{\mathbf{k}}}\rightarrow e^{i\varphi^{v}(\mathbf{k})}\ket{u^{v}_{\mathbf{k}}}$, $F(\mathbf{Q})\rightarrow e^{i(\varphi^{v}(-\mathbf{Q}/2))-\varphi^{c}(\mathbf{Q}/2)}e^{i\theta(\mathbf{Q})}F(\mathbf{Q})$ and $\ket{U^{exc}(\mathbf{Q})}\rightarrow e^{i\theta(\mathbf{Q})}\ket{U^{exc}(\mathbf{Q})}$, we can verify that the section is gauge invariant and, hence, globally defined. Notably, the gauge transformation of $F(\mathbf{Q})$  matches that which renders the Berry curvature of the exciton in Eq.~\eqref{eq:Berry_Connection_Exciton} a gauge invariant quantity~\cite{Kwan_2021}. Notice that, by construction, in the case of no interactions $\mathcal{L}_{F}$ is a trivial line bundle, hence the sections $F(\mathbf{Q})$ are constant functions. The construction then reduces to $\mathcal{L}_{int}=\mathcal{L}_{0}$, which is the well-known case of non-interacting particles.

We can, then, establish a singular connection in the complex line bundle $\mathcal{L}_{F}\rightarrow BZ^2$ using this global section. Besides the singular connection, $\mathcal{L}_{F}$ also inherits a naturally induced connection from the connections defined on $\mathcal{L}_{0}$ and $\mathcal{L}_{int}$, given by Eq.~\eqref{eq:Berry_Connection_L0} and Eq.~\eqref{eq:Berry_Connection_Exciton}, respectively:
\begin{equation}
\mathcal{A}^{F} = (\mathcal{A}^{c}-\mathcal{A}^{v})-\mathcal{A}^{exc}.
\label{eq:Connection_"Berry"_F}
\end{equation}
This connection is significant as it quantifies, in geometric terms, the difference between the exciton bundle and the independent particle bundles. Specifically, if the line bundle $\mathcal{L}_{F}$ is trivial, then the two bundles from which we construct $\mathcal{L}_{F}$, $\mathcal{L}_{0}$ and $\mathcal{L}_{int}$, are isomorphic and thus topologically equivalent. Physically, this connection measures the polarization of the exciton system relative to the creation and annihilation of the independent particles, hole, and electron, with respect to the unit cell. Therefore, this is a gauge-dependent quantity. However, because the three connections, $\mathcal{A}^{c}$, $\mathcal{A}^{v}$, and $\mathcal{A}^{exc}$, are related, it must be possible to construct a gauge-independent shift. This is the motivation for constructing the singular connection, as it is the missing piece needed to create a gauge-invariant object for the exciton.

Moreover, given that we have a global section of $\mathcal{L}_{F}$ solely determined by the coefficients $F(\mathbf{Q})$, we can define a singular connection on it by imposing equation~\eqref{eq:singular_connection_def} on the chosen global section of $\mathcal{L}_{F}$. This approach yields the connection 1-form of the singular connection:

\begin{equation}
\mathcal{A}^{F}_{sing} = i\mathrm{d}\log{(F(\mathbf{Q}))}.
\label{eq:Connection_Singular_F}
\end{equation}
It is evident from the obtained expression for the singular connection in Eq.~\eqref{eq:Connection_Singular_F} that it encapsulates information solely about the coefficients $F(\mathbf{Q})$. We will now write down the singular curvature in Eq.\eqref{eq:singular_curvature_general} for the exciton case. Analogously to what was done in Eq.\eqref{eq:singular_curvature_general} we express the coefficients $F(\mathbf{Q})=x_{1}(\mathbf{Q})+ix_{2}(\mathbf{Q})$ in terms of its real and imaginary part and obtain:

\begin{equation}
    \Omega^{F}_{sing} = -2\pi \delta^{2}(F(\mathbf{Q}))\mathrm{d}x_{1}(\mathbf{Q})\wedge \mathrm{d}x_{2}(\mathbf{Q}).
    \label{eq:singular_cunvature_F}
\end{equation}
In this scenario, the curvature becomes infinite at points in the Brillouin zone $BZ^2$ where the global section vanishes, corresponding to the zeros of the interaction coefficients $F(\mathbf{Q})$. For a finite number of such generic zeros, Eq.\eqref{eq:singular_cunvature_F} can be more intuitively expressed as:

\begin{equation}
\Omega^{F}_{sing}=-2\pi \sum_{\mathbf{Q}^{*}\in N_V} \sigma_{\mathbf{Q}^{*}}\delta^{2}(\mathbf{Q}-\mathbf{Q}^{*})\mathrm{d}\mathbf{Q}_{x}\wedge \mathrm{d}\mathbf{Q}_{y},
\label{eq:OmegaSingularZeros}
\end{equation}
where $\sigma_{\mathbf{Q}^{*}}\in \{\pm 1\}$ indicates whether the map $F(\mathbf{Q})$ is orientation-preserving or -reversing in the vicinity of $\mathbf{Q}^*\in BZ^2 \text{ such that }F(\mathbf{Q^*})=0$. We pause here to make the connection between the result derived here and the one appearing in Ref.~\cite{Mera_2021} where the singular connection is used to study the transition dipole matrix element for optical transitions between the two bands for linearly polarized light. In Ref.~\cite{Mera_2021}, the singular curvature is a measure for the number of points in the Brillouin zone where transitions between the two bands being considered are forbidden. In our work a similiar intuition can be made in order to interpret Eq.~\ref{eq:OmegaSingularZeros}. The singular curvature $\Omega^{F}_{sing}$ in Eq.~\eqref{eq:OmegaSingularZeros} counts the number of points in the Brillouin zone of the exciton for which it is not possible for the electron and hole to bind and form an exciton. That is, under the effect of the interaction potential and in combination with the band structure, the information about the possibility for binding excitons is specified by the interaction coefficients $F(\mathbf{Q})$.

The singular structure of the curvature can also be thought of as analogous to the vorticity of the zeros of electronic wavefunctions of electrons on the lattice in the presence of an external magnetic field, as described in~\cite{kohmoto}. The parallelism between the Chern number, traditionally obtained by integrating the Berry curvature, and counting the vorticities of the zeros of an electronic wavefunction introduced in~\cite{kohmoto} is made more explicit by using the singular curvature. By integrating $\Omega^{sing}_{F}$ over the entire Brillouin zone, the topological invariant is determined, counting the points in the Brillouin zone, including their vorticity i.e., the sign of the map $\sigma_{\mathbf{Q}^*}$, where the interaction coefficients $F(\mathbf{Q})$ are zero. In essence, it highlights the singularities themselves. Furthermore, since the integral of the singular curvature, as explained before, yields the Chern number, the bridge established in~\cite{kohmoto} between the Chern number and the vorticities of the wavefunction's zeros is essentially encapsulated in the concept of singular curvature. This provides a direct and intuitive understanding of the topological invariant in terms of the zeros and their vorticities. Note that if we had chosen a different value of $\mathbf{p}$, it could happen that the number of zeros of the interaction coefficients $F(\mathbf{Q})$ would differ from those at $\mathbf{p} = 0$. However, the number of zeros weighted by their vorticity would match those for $\mathbf{p} = 0$, hence the chern number remains the same. 

To prepare for the following sections, it's important to understand how to reintroduce $\mathbf{p}$ into the geometric quantities defined earlier. By selecting a generic value for $\mathbf{p}$, we obtain a different homomorphism bundle $\mathcal{L}_{F}$, which is isomorphic to the one constructed with $\mathbf{p} = 0$. While these bundles share the same topological content due to their isomorphism, the local geometric quantities differ. For the single-particle Berry connections, a generic $\mathbf{p}$ value results in a translation in momentum space within the single particle BZ$^2$, where the valence and conduction band connections, denoted by $\mathcal{A}^{v}(\mathbf{p})$ and $\mathcal{A}^{c}(\mathbf{p})$, are shifted in opposite directions. The global section used to define the singular curvature also changes and is represented by $F(\mathbf{Q},\mathbf{p})$, and the singular connection for a generic $\mathbf{p}$ is denoted by $\mathcal{A}^{F}_{sing}(\mathbf{p})$. Introducing a finite $p$ changes the local singular structure but leaves the total vorticity unchanged.

\section{Exciton Shift Vector  \texorpdfstring{$\mathbf{S}_{\mathbf{Q}}$}{SQ}}
\label{sec:Shift_Vector}

We now proceed similarly to~\cite{Mera_2021}, constructing a shift vector derived from the disparity between an induced connection on the homomorphism bundle and the singular connection on that same bundle. The shift vector holds significance not only due to its gauge invariance, making it measurable, but also because of its tangible interpretation~\cite{vonBaltz1981,Pesin_2018,Kaplan2022}: measuring the displacement of the particle's center of charge following its transition to a different energy band. 
Namely, when a particle-hole pair is created, the associated shift of the charge center in a material that breaks inversion symmetry can be quantified by the single-particle shift vector $\mathbf{S}_{cv}$, defined as
\begin{align}
    \mathbf{S}_{cv}
    = \mathcal{A}^{c}-\mathcal{A}^{v}-i\mathrm{d}\log{\mathcal{F}},
\end{align}
where $\mathcal{F}$ is the single-particle interband Berry connection, which is non-Abelian. Here $i\mathrm{d}\log\mathcal{F}$ constitutes the singular connection between valence and conduction band~\cite{Mera_2021}.

In the context of the exciton, the same reasoning translates to a shift in the center of mass of the electron-hole pair upon the formation of the bound state. 
Based on our results of the previous section we therefore propose to define an exciton's shift vector for each $\mathbf{p}$ as follows:
\begin{align}
\mathbf{S}_{\mathbf{Q},\mathbf{p}} &:= \mathcal{A}^{F}(\mathbf{p})-\mathcal{A}^{F}_{sing}(\mathbf{p}) 
\notag\\ &
= \mathcal{A}^{cv}(\mathbf{p})-\mathcal{A}^{exc}-i\mathrm{d}\log{F(\mathbf{Q},\mathbf{p})}.
\label{eq:shift_vector_p}
\end{align}
As pointed out before, for the topology of the exciton the dependence in $\mathbf{p}$ is irrelevant. Hence to discuss the topology of the exciton and how it is expresses itself in the shift vector, we will now first focus on the shift vector for $\mathbf{p}=0$:
\begin{equation}
\mathbf{S}_{\mathbf{Q}} := \mathbf{S}_{\mathbf{Q},\mathbf{p}=0} = (\mathcal{A}^{c}-\mathcal{A}^{v})-\mathcal{A}^{exc}-i\mathrm{d}\log{F(\mathbf{Q})},
\label{eq:shift_vector_p0}
\end{equation}
Given that this is the difference between two connections within the same line bundle, it qualifies as a gauge-invariant quantity, thus a globally defined 1-form. As elucidated earlier, the 2-form $\mathrm{d}\mathbf{S}_{\mathbf{Q}}$ is exact, therefore its integration over the entire Brillouin zone $BZ^2$ vanishes. Alternatively, the gauge invariance of the shift vector can be viewed in the following way. In the last section, we discussed that $\mathcal{A}^{F}$ is a gauge-dependent quantity. This is because $\mathcal{A}^{exc}$ includes not only the phases of the single particles (electron and hole) but also the phases of the interaction coefficients $F(\mathbf{Q})$, which are not considered in the single particle connections $\mathcal{A}^{c}$ and $\mathcal{A}^{v}$. The singular connection, $\mathcal{A}^{F}_{sing}$, contains only the phases of the interaction coefficients $F(\mathbf{Q})$. Thus, the singular connection corrects the gauge dependence of $\mathcal{A}^{F}$, leading to a gauge-invariant object for the exciton, which is the shift vector. 

Physically, we can then interpret Eq.~\eqref{eq:shift_vector_p0} as follows: 
The phase accumulation of the exciton along a trajectory in momentum space, determined by $\mathcal{A}^{exc}$, has four different contributions. Two contributions come from the phase shift of the single particles, electron and hole, determined by $\mathcal{A}^{c}$ and $\mathcal{A}^{v}$, respectively. Another one comes from the phase accumulation due to the interaction coefficients $F(\mathbf{Q})$. The final contribution comes from the shift vector, which can be interpreted as the difference in the center of mass between the exciton and the original center of mass of the single particles, electron and hole, with respect to the unit cell.
We argue that the shift vector suffices to elucidate the distinctions in the topology of the exciton compared to that of the independent particles.

After defining the topological content of the exciton including the changes induced by Coulomb interaction on the electron and hole by means of the shift vector $\mathbf{S}_{\mathbf{Q}}$, we discuss its characteristics in several limiting cases. 
We start by examining how the topology of independent particles is reinstated in the weak coupling regime, in which the interaction between the two particles can be neglected. Subsequently, we investigate how the interaction coefficients introduce new topological features and their correlation with changes in the system polarization.

In the weak coupling regime, one returns to the scenario where the electron and the hole behave as independent entities. This state is characterized by uniform, non-zero, coefficients $F(\mathbf{Q})$. Consequently, the singular connection segment of the shift vector in Eq.~\eqref{eq:shift_vector_p0} vanishes, simplifying the expression to:
\begin{equation}
\mathbf{S}_{\mathbf{Q}} = (\mathcal{A}^{c}-\mathcal{A}^{v})-\mathcal{A}^{exc}.
\end{equation}
Moreover, as previously discussed, shifting focus to curvatures rather than connection 1-forms reveals that $\mathrm{d}\mathbf{S}_{\mathbf{Q}}$ is an exact two-form. This implies that the Berry curvature of the exciton and the difference in Berry curvatures of individual particles is zero at the cohomology level, i.e. $[\Omega^{exc}_{Berry}]=[\Omega^{c}_{Berry}-\Omega^{v}_{Berry}]$, where $\Omega^{exc}_{Berry}=\mathrm{d}\mathcal{A}^{exc}$, $\Omega^{c}_{Berry}=\mathrm{d}\mathcal{A}^{c}$, and $\Omega^{v}_{Berry}=\mathrm{d}\mathcal{A}^{v}$. Consequently, upon integration across the Brillouin zone, the Chern number of the exciton band aligns with the difference in Chern numbers of conduction and valence bands. We will refer to these as trivial excitons.

As the coefficients $F(\mathbf{Q})$ deviate increasingly from uniformity, their possible impact on the topology of independent particles becomes non-negligible, altering the exciton's topology from that of the independent particles. By examining curvatures, as previously noted, the 2-form $\mathrm{d}\mathbf{S}_{\mathbf{Q}}$ emerges as an exact two-form. Consequently, the disparity between the Berry curvature of the exciton and that of the single particles, along with the curvature of the singular connection, falls within the same cohomology class: $[(\Omega^{c}_{Berry}-\Omega^{v}_{Berry})-\Omega^{exc}_{Berry}]=[\Omega_{sing}]$. The nontrivial nature of the singular curvature, detailed in Eq.\eqref{eq:singular_cunvature_F}, imparts distinct topology to the exciton band, makingits Chern number deviate from the difference in Chern numbers of conduction and valence bands. This deviation hinges solely upon the coefficients representing particle interaction, $F(\mathbf{Q})$. We will refer to these excitons as topological.

Thus, we discern that the nontrivial topology introduced in the exciton's problem due to interaction resides in the coefficients $F(\mathbf{Q})$, or geometrically, in the singular curvature denoted by Eq.\eqref{eq:singular_cunvature_F}. In fact, we can make this deviation in the Chern numbers more intuitive in the scenario in which the coefficients $F(\mathbf{Q})$ have a finite number of zeros in the Brillouin zone. Based on the discussion in Sec.~\ref{sec:singular_connection}, we conclude that the Chern number of the exciton is given by:
\begin{equation}
C^{exc} = C^{c}-C^{v}-\bar{N}_V,
\end{equation}
with $\bar{N}_V$ being the number of zeros of $F(\mathbf{Q})$ in the Brillouin zone weighted by their vorticity. It can happen that $\bar{N}_V$ is zero, even though $F(\mathbf{Q})=0$ for some $\mathbf{Q}\in$BZ$^2$ since its vorticities, when summed, might cancel each other. 
However, it is a necessary condition that for an exciton to be topological, i.e. for its Chern number to differ from the one of the trivial exciton, the coefficients $F(\mathbf{Q})$ must vanish for some points in the Brillouin zone. Furthermore using Eq.~\eqref{eq:OmegaSingularZeros}:
\begin{equation}
|\bar{N}_V|=\Big|\sum_{\mathbf{Q^{*}}\in N_V}\sigma
_{\mathbf{Q^{*}}}\Big|\leq \sum_{\mathbf{Q^{*}}\in N_V} 1 = \text{ no.~of zeros of $F(\mathbf{Q})$}.
\end{equation}
This means that when the Chern number of the singular curvature is $\bar{N}_V$ we expect a minimum of $|\bar{N}_V|$ zeros of the coefficients $F(\mathbf{Q})$.

\begin{figure*}[t]
\centering
\includegraphics[width=.85\textwidth]{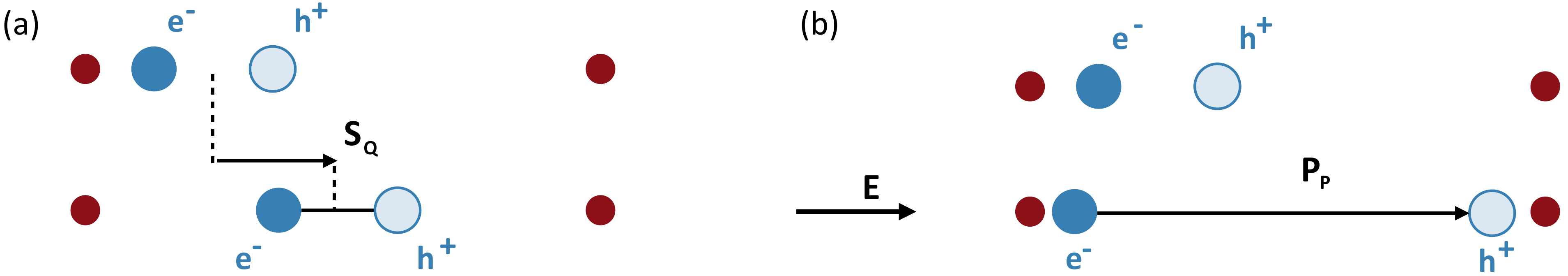}
\caption{
\label{fig:polarizations}
{\small
Schematic representation of the two polarization vectors that characterize the combined position and spread of the electron (blue) and hole (white) with respect to the lattice (red). 
(a)  The top picture depicts the positions of electron and hole in the non-interacting limit before exciton formation, while the bottom picture depicts the exciton as the bound state of the electron and the hole. The center of mass displacement that occurs when the exciton is formed is encapsulated by the polarization vector $\mathbf{S}_{\mathbf{Q}}$. One can interpret this polarization, as an accumulation of adiabatic phases that have three distinct origins; The conduction and valence band and the winding of the coefficients $F(\mathbf{Q})$ around its zeros all contribute to the adiabatic phases. 
(b) In the presence of an external electric field, the electron and the hole that constitute the exciton respond differently to it, which results in a change of the total energy in presence of the field. This is captured by the polarization vector $\mathbf{P}_{\mathbf{p}}$.
}
}
\end{figure*}

Moreover, we can draw a parallel between a nontrivial Chern number in an insulator and a nontrivial polarization, attributing a physical significance to this new Chern number, coming from the coefficients $F(\mathbf{Q})$, in the exciton's topology. Should the singular curvature prove nontrivial, the exciton wavefunction gains an additional positional shift owing to interactions, yielding nontrivial excitons. Indeed, the shift vector can be interpreted as a polarization vector which contains information about the adiabatic phases gained upon the formation of the exciton. These phases have three distinct origins; from the conduction and valence band and from the winding of the coefficients $F(\mathbf{Q})$ around its zeros. These phases all contribute to the polarization vector. Trivial excitons will not have the latter contribution. The shift vector interpreted as a polarization vector is illustrated in a 1D lattice for better visualization in Fig.\ref{fig:polarizations} (a). Importantly, the shift vector does not measure the distance between the electron and hole that constitute the exciton. Such an inter-particle distance is an internal degree of freedom for the exciton which is measured by another polarization vector, to be discussed next.

\section{Polarization \texorpdfstring{$\mathbf{P}_{\mathbf{p}}$}{Pp}}
\label{sec:relative_polarization}
As explained in the previous section, for the topology of the exciton the relative momentum  $\mathbf{p}$ cannot play a role. 
Nonetheless, since the relative coordinate $\mathbf{r}$ between the electron and hole constitutes the polarization vector of the exciton, it is expected to be relevant for certain dynamical properties. 
This notion can be made precise by computing the expectation value of the relative coordinate operator $\mathbf{\hat{r}}$ with respect to a normalized exciton wavepacket:
\begin{equation}
  \Psi^{exc}_{\omega} = \sum_{\mathbf{Q}}\omega(\mathbf{Q}) \psi^{exc}_{\mathbf{Q}},
    \label{eq:exciton_wavepacket}
\end{equation}
$\omega(\mathbf{Q})$ is a Gaussian envelope centered around $\mathbf{Q}=\mathbf{Q_\omega}$. This envelope is normalized to ensure the normalization of $\Psi^{exc}_\omega$:
\begin{equation}
   \sum_{\mathbf{Q}} |\omega(\mathbf{Q})|^2 = 1.
    \label{Normalization}
\end{equation}
The expectation value $\mathbf{r}_\omega=\bra{\Psi^{exc}_\omega}\hat{\mathbf{r}}\ket{\Psi^{exc}_\omega}$ then becomes well defined, and can be written as (cf. App. C)
\begin{align}
\mathbf{r}_\omega(\mathbf{Q}_\omega) &= 
\sum_{\mathbf{p}}|F(\mathbf{p},\mathbf{Q_\omega})|^2 
\Bigl(
\mathcal{A}^{c}\bigl(\mathbf{p}+\tfrac{\mathbf{Q}_\omega}{2}\bigr) - 
\mathcal{A}^{v}\bigl(\mathbf{p}-\tfrac{\mathbf{Q}_\omega}{2}\bigr)\Bigr)
\nonumber\\
&\qquad+i\sum_{\mathbf{p}} F^{*}(\mathbf{p},\mathbf{Q_\omega})\mathbf{\nabla}_{\mathbf{p}} F(\mathbf{p},\mathbf{Q_\omega})\nonumber\\
&:= \sum_{\mathbf{p}}\mathbf{P}_{p}(\mathbf{Q}_\omega).
\label{eq:P_p}
\end{align}
Notably, the quantity $\mathbf{P}_{p}$ itself is gauge invariant (cf. App B for a proof), strongly suggesting that it constitutes the intrinsic  physically observable polarization of the exciton. We point out that a polarization termed ``quantum geometric dipole" which enters into the equations of motion was previously constructed in~\cite{Fertig_2021}, which contained the interaction coefficients implicitly. It is not obvious if the two quantities represent the same object, as the underlying assumptions of both model are different.

\section{Semiclassical Equations of motion}
\label{sec:semiclassical_equations_motion}

In this section, we establish the wave packet dynamics of excitons under the influence of an electric field and highlight the natural emergence of the  polarization vector $\mathbf{P}_{\mathbf{p}}$.
A few studies have addressed the semiclassical equations of motion for excitons~\cite{Niu_2008,Fertig_2021,Chaudhary_2021}. 
However, it is desirable to formulate the semiclassical motion of an exciton in a way which makes explicit both the effects of the wavefunction renormalization $F(\mathbf{p},\mathbf{Q})\neq1$ and of the exciton's internal degrees of freedom.
Namely, in Ref.~\cite{Niu_2008} the effective Lagrangian is treating the exciton as a single particle (cf. also~\cite{Niu_1995}). The resulting equations mirror those of a single particle, with the exciton's position, momentum, and Berry curvature replacing those of the individual particles. Conversely, in Ref.~\cite{Fertig_2021}, the exciton is treated as an electron-hole pair connected by a ``rod", where the role of the interaction coefficient is only implicit. 

Our aim is to elucidate all effects of the coefficients $F(\mathbf{p},\mathbf{Q})$ in the equations of motion. 
Since finding the phase introduced in the exciton's wavefunction by the minimal coupling formalism can be challenging, we pick 
an alternative strategy that involves adding to the two-particle Hamiltonian a dipole term: $e\mathbf{E}\cdot \hat{\mathbf{r}}$, where $e$ is the electron charge and $\mathbf{E}$ denotes the external electric field, which couples with the operator of the exciton's relative coordinate $\hat{\mathbf{r}}$. 
To compute the semiclassical equations of motion of the exciton, we follow the procedure outlined in~\cite{Niu_1995}, starting with the computation of the effective Lagrangian of the system,
\begin{equation}
\mathcal{L} = \bra{\Psi^{exc}_\omega}i\mathrm{d}/\mathrm{d}t\ket{\Psi^{exc}_\omega}-\bra{\Psi^{exc}_\omega}\Delta H\ket{\Psi^{exc}_\omega}-\epsilon^{exc}_{\mathbf{Q}},
\label{eq:effective_Lagrangian}
\end{equation}
with $\Delta H = e\mathbf{E}\cdot \hat{\mathbf{r}}$, $\epsilon^{exc}$ the exciton energy that can be obtained from the Bethe Salpeter equation Eq.~\eqref{eq:BSE} and $\Psi^{exc}_\omega$ given by Eq.~\eqref{eq:exciton_wavepacket}.
The piece of the effective Lagrangian associated to $\Delta H$ evaluates directly to $e\mathbf{E}\cdot \mathbf{r}_\omega(\mathbf{Q}_\omega)$, with $\mathbf{r}_\omega$ defined in Eq.~\eqref{eq:P_p} in terms of the polarization vector $\mathbf{P}_{\mathbf{p}}$.

Next, we compute the first term in the effective Lagrangian:
\begin{equation}
\bra{\Psi^{exc}_\omega}i\mathrm{d}/\mathrm{d}t\ket{\Psi^{exc}_\omega} = \mathbf{Q}_\omega\cdot \mathbf{\Dot{R}}_\omega + \mathbf{\Dot{Q}}_\omega\cdot \mathcal{A}^{exc}(\mathbf{Q}_\omega).
\label{eq:time_dependance_effective_Langrangean}
\end{equation}
Here, the exciton Berry connection is:
\begin{align}
\mathcal{A}^{exc}(\mathbf{Q})
&= \sum_{p} \frac{|F(\mathbf{p},\mathbf{Q})|^2}{2} \Bigl(
\mathcal{A}^{c}\bigl(\mathbf{p}+\tfrac{\mathbf{Q}}{2}\bigr) + 
\mathcal{A}^{v}\bigl(\mathbf{p}-\tfrac{\mathbf{Q}}{2}\bigr)\Bigr)
\nonumber\\&\quad
+  i \sum_{p} F^{*}(\mathbf{p},\mathbf{Q})\mathbf{\nabla}_{\mathbf{Q}} F(\mathbf{p},\mathbf{Q}) ,
\end{align}
which closely resembles the exciton connection first derived in~\cite{Niu_2008} within the effective mass approximation.
Combining all terms, we thus arrive at the effective Lagrangian:
\begin{align}
\mathcal{L} &= \mathbf{Q}_\omega\cdot \mathbf{\Dot{R}}_\omega + \mathbf{\Dot{Q}}_\omega\cdot \mathcal{A}^{exc}(\mathbf{Q}_\omega)
-e\mathbf{E}\cdot\mathbf{r}_\omega(\mathbf{Q}_\omega) -\epsilon^{exc}_{\mathbf{Q}_\omega}.
\end{align}

We explicitly compute the exciton's Berry curvature using the standard definition $\Omega^{exc}(\mathbf{Q}):=\Omega^{exc}_{Berry}(\mathbf{Q})=\mathrm{d}\mathcal{A}^{exc}_{Berry}$ (Eq.~\eqref{eq:Berry_Connection_Exciton}) and inserting the Bloch wavefunctions $U^{exc}(\mathbf{Q})$ of the exciton (Eq.~\ref{eq:exciton_wavefunction}). We obtain
\begin{align}
&\Omega^{exc}(\mathbf{Q}) = -i\sum_{\mathbf{p}}\Big(\frac{\partial F^{*}}{\partial Q_x}\frac{\partial F}{\partial Q_y}-\frac{\partial F^{*}}{\partial Q_y}\frac{\partial F}{\partial Q_x}\Big) +\nonumber\\
&+ \frac{1}{2}\sum_{\mathbf{p}} \Big(\frac{\partial |F(\mathbf{p},\mathbf{Q})|^2}{\partial Q_x} \mathcal{A}^{c}_{y}(\mathbf{p}+\mathbf{Q}/2)- x\leftrightarrow y \Big)\nonumber\\
&+ \frac{1}{2}\sum_{\mathbf{p}} \Big(\frac{\partial |F(\mathbf{p},\mathbf{Q})|^2}{\partial Q_x} \mathcal{A}^{v}_{y}(\mathbf{p}-\mathbf{Q}/2) - x\leftrightarrow y \Big)\nonumber\\
\label{eq:Berry_Curvature_Exciton}
&+ \frac{1}{4}\sum_{\mathbf{p}} |F(\mathbf{p},\mathbf{Q})|^2 \Omega^{c}(\mathbf{p}+\mathbf{Q}/2) \nonumber\\
&- \frac{1}{4}\sum_{\mathbf{p}} |F(\mathbf{p},\mathbf{Q})|^2 \Omega^{v}(\mathbf{p}-\mathbf{Q}/2),
\end{align}
with $\Omega^{v/c}$ the Berry curvatures of the valence and conduction bands.
Although the final expressions appearing here closely resemble those derived in ~\cite{Niu_2008,Kwan_2021} within the effective mass approximation, the approach presented here is drastically different and applies across the entire BZ$^2$ and for any set of isolated bands. 
The somewhat unconventional form of the exciton Berry curvature arises due to non-uniform coefficients $F(\mathbf{p},\mathbf{Q})$. Although the explicit appearance of the single-particle Berry connections in $\Omega^{exc}$ may raise concerns regarding gauge invariance, we have verified that it is ~\cite{Kwan_2021}.

We are now in the position to derive the semiclassical equations of motion,
\begin{align}
\label{eq:Rdot}
\Dot{\mathbf{R}}_{\omega} &= 
\mathbf{\nabla}\!_{\mathbf{Q}_{\omega}} \epsilon^{exc}_{\mathbf{Q}_\omega}-\Dot{{\mathbf{Q}}}_{\omega}\times\Omega^{exc}(\mathbf{Q}_\omega)\nonumber\\
&-e\mathbf{\nabla}\!_{\mathbf{Q}_{\omega}}\sum_{\mathbf{p}}\bigl(\mathbf{E}\cdot \mathbf{P}_{p}(\mathbf{Q}_\omega)\bigr)\\
\label{eq:Qdot}
\Dot{\mathbf{Q}}_{\omega} &= -e\mathbf{\nabla}_{\mathbf{R}_{\omega}} \mathbf{E} \cdot\Big(\sum_{\mathbf{p}}\mathbf{P}_{p}(\mathbf{Q}_\omega)\Big).
\end{align}
One can make the relation with the electronic band structure quantities explicit by using the $\mathbf{p}$ dependent shift vector $\mathbf{S}_{\mathbf{Q},\mathbf{p}}$ in Eq.~\eqref{eq:shift_vector_p}. In particular, by taking the exterior derivative of the connections and using the fact that in 2d $d\mathbf{S}_{\mathbf{Q}}=\nabla_{\mathbf{Q}}\times\mathbf{S}_{\mathbf{Q}}$. Then we have that:
\begin{align*}
\Omega^{exc}(\mathbf{Q}_{\omega}) &= \sum_{\mathbf{p}}(\Omega^{c}(\mathbf{Q}_{\omega},\mathbf{p})-\Omega^{v}(\mathbf{Q}_{\omega},\mathbf{p})-\Omega_{sing}(\mathbf{Q}_{\omega},\mathbf{p})\\
&-\nabla_{\mathbf{Q}}\times\mathbf{S}_{\mathbf{Q}_\omega,\mathbf{p}}).
\end{align*}
Using this equality we obtain:
\begin{align}
\label{eq:RdotShift}
\Dot{\mathbf{R}}_{\omega} &= 
\mathbf{\nabla}\!_{\mathbf{Q}_{\omega}} \epsilon^{exc}_{\mathbf{Q}_\omega}-\Dot{{\mathbf{Q}}}_{\omega}\times\Big(\sum_{\mathbf{p}}\Omega^{c}(\mathbf{Q}_\omega,\mathbf{p})-\Omega^{v}(\mathbf{Q}_\omega,\mathbf{p})\nonumber\\
&-\Omega_{sing}(\mathbf{Q}_\omega,\mathbf{p})\Big)+\Dot{\mathbf{Q}}_{\omega}\times \nabla_{\mathbf{Q}}\times \sum_{\mathbf{p}}\mathbf{S}_{\mathbf{Q},\mathbf{p}}\nonumber\\
&-e\mathbf{\nabla}\!_{\mathbf{Q}_{\omega}}\sum_{\mathbf{p}}\bigl(\mathbf{E}\cdot \mathbf{P}_{p}(\mathbf{Q}_\omega)\bigr)\\
\label{eq:QdotShift}
\Dot{\mathbf{Q}}_{\omega} &= -e\mathbf{\nabla}_{\mathbf{R}_{\omega}} \mathbf{E} \cdot\Big(\sum_{\mathbf{p}}\mathbf{P}_{p}(\mathbf{Q}_\omega)\Big).
\end{align}

From the last term in the equation of motion Eq.~\eqref{eq:Rdot}, one can read of the physical interpretation of $\mathbf{P}_{\mathbf{p}}$. This term is added to the anomalous velocity alongside the well-known Berry curvature term. Notably, unlike the latter, this new anomalous velocity can have a longitudinal component.
Furthermore, it couples to the electric field in the form of a dipole energy which renormalizes the dispersion, reflecting the fact that even though the exciton is a charge neutral quasi-particle, it is still made up of an electron and a hole, and so it has an intrinsic electric dipole that allows it to couple to the electric field (Fig.~\ref{fig:polarizations}). Finally, we would like to point out that for the $\mathbf{Q}$ points in BZ$^2$ for which $\Omega_{sing}$ has a singular structure this one is compensated by the singular structure in $\mathbf{S}_{\mathbf{Q},\mathbf{p}}$, so there isn't overall any singular structure in the semiclassical equations of motion. However, if there are band touching points in the exciton bands then the semiclassical equations of motion are not defined in those points.

In summary, the semiclassical equations of motion Eqs.~(\ref{eq:Rdot},\ref{eq:Qdot}), together with the band-periodic definitions Eqs.~(\ref{eq:P_p},\ref{eq:Berry_Curvature_Exciton}) present a complete formulation of the semiclassical effects of the quantum geometry of excitons, including all effects of the electron-hole interaction on the wavefunctions. 
In contradistinction, the more verbose formulation of Eq.~\eqref{eq:RdotShift} makes it obvious that the exciton properties are determined by three different ingredients, the electronic band topology, the vorticity of the interaction coefficients and finally the shift of the center of mass.

\section{Conclusion and Discussion}
\label{sec:conclusion}
In this work, we have derived and characterized the geometric and topological properties of excitonic states by utilizing the full lattice wavefunctions, bypassing the limitations of the effective mass approximation. Our work provides a comprehensive framework for understanding the role of the interaction coefficients $F(\mathbf{p},\mathbf{Q})$ regarding the charge distribution of exciton states, addressing a substantial gap in current knowledge.
We find that the the singular connection plays a crucial role in determining the properties of the exciton wavefunction. We show that this new connection is defined solely by the interaction coefficients of the exciton, measuring their vortex structure in momentum space.

We have identified two key gauge-invariant quantities that fully characterize the exciton's topology and geometry: the exciton shift vector $\mathbf{S}_{\mathbf{Q}}$ and the exciton dipole vector $\mathbf{P}_{\mathbf{p}}$. The shift vector $\mathbf{S}_{\mathbf{Q}}$ encapsulates the influence of electron-hole interactions on the exciton band topology.
By studying the shift vector, we identified under which circumstances the topology of the exciton bands differs from the parent electronic band structure, finding that interactions can introduce nontrivial topology in the exciton bands beyond the topology which is contained in the single-particle band structure.
The shift vector also reveals a nontrivial polarization that may be observable at elevated temperatures with finite exciton densities. The dipole vector $\mathbf{P}_{\mathbf{p}}$ is associated with the exciton's intrinsic dipole moment and plays a crucial role in exciton dynamics, leading to an anomalous velocity term in the exciton's semiclassical equations of motion.

Despite these advancements, several avenues for future research remain. Numerical calculations to obtain expressions for the interaction coefficients $F(\mathbf{p},\mathbf{Q})$, along with density functional theory (DFT) calculations to determine the topology of materials hosting excitons, can shed further light on the role of interactions in exciton topology. 
We also believe that it is possible with current devices to probe the  the exciton shift vector and the exciton dipole vector, for example by measurement of transport and inverse compressibility as a function of temperature. 
Investigating the impact of these geometric and topological properties on other exciton characteristics, such as optical transitions, presents a promising direction for further study. 

\begin{acknowledgments}
C.P. is grateful to Bruno Mera for very insightful and fruitful discussions and for carefully reading parts of the manuscript. 
The authors would like to thank Sivan Refaely-Abramson, Herb Fertig and Bruno Uchoa for illuminating discussions.  
T.H. acknowledges financial support by the 
European Research Council (ERC) under grant QuantumCUSP
(Grant Agreement No. 101077020). C.P. And R.I. were supported by the US-Israel Binational Science Foundation (BSF) Grant No.2018226.
\end{acknowledgments}

\section*{Appendix A: Independence of Coordinate Basis of Berry Curvature}

In this appendix we prove that the Berry curvature, the 2-form object, is independent of the coordinate basis we choose to represent it in. The proof will be done for a generic manifold $B$, which will be parameterized by the local coordinates $\mathbf{k}=\{k_{i}\}$ with $i=1,...,d$ and $d$ is the dimension of the manifold $B$. In our case the manifold $B$ is the first Brillouin zone of the exciton (BZ$^{2}$), which is diffeomorphic to a torus $\mathbb{T}^{2}$. The Berry curvature can then be written using the local coodinates $k_{i}$ in the following way:
\begin{align}
    \Omega = \frac{1}{2}\sum_{i,j} \Omega_{ij}(\mathbf{k})\mathrm{d}k_i\wedge \mathrm{d}k_j .
\end{align}
We now perform a change of basis in the local coordinates $k_{i}$, such that $k_{i}=k_{i}(x_1,...,x_d)$. There is a change from the local coordinates $k_{i}$ to the local coordinates $x_{i}$. This change in local basis reflects itself in the following way in the Berry curvature:
\begin{align}
\Omega &= \frac{1}{2}\sum_{i,j} \Omega_{ij}(\mathbf{k})\mathrm{d}k_i\wedge \mathrm{d}k_j \nonumber\\
&= \frac{1}{2}\sum_{i,j}\sum_{m,n} \Omega_{ij}(\mathbf{k(\mathbf{x})})\frac{\partial k_i}{\partial x_m} \frac{\partial k_j}{\partial x_n} \mathrm{d}x_m\wedge \mathrm{d}x_n\nonumber\\
&= \frac{1}{2}\sum_{m,n} \overline{\Omega}_{mn}(\mathbf{x})\mathrm{d}x_m\wedge \mathrm{d}x_n,
\end{align}
with $\overline{\Omega}_{mn}(\mathbf{x})=\sum_{ij}\Omega_{ij}(\mathbf{k(\mathbf{x})})\frac{\partial k_i}{\partial x_m} \frac{\partial k_j}{\partial x_n}$.
From the last line of the computation above we are able to conclude that the Berry curvature (the 2-form) is independent of the choice of local coordinates. 
Consequently, observables that are topological invariants (Chern numbers), i.e. that result from the integration of the Berry curvature in the manifold $B$ are also naturally independent of the choice of local coordinates. 
\vspace{2mm}
\section*{Appendix B: Gauge Invariance of \texorpdfstring{$\mathbf{P}_{\mathbf{p}}$}{Pp}}

In this appendix, we establish the gauge invariance of the newly defined quantity $\mathbf{P}_{\mathbf{p}}$ as introduced in the main text. To accomplish this, we operate under a gauge transformation that renders the Berry curvature of the exciton, as given in Eq.~\eqref{eq:Berry_Curvature_Exciton}, a gauge-invariant quantity. This transformation mirrors that introduced in ~\cite{Kwan_2021}:
\begin{align}
\label{eq:gauge_uc}
&\ket{u^{c}_{\mathbf{k}}}\rightarrow e^{i\varphi^{c}(\mathbf{k})}\ket{u^{c}_{\mathbf{k}}}\\
\label{eq:gauge_uv}
&\ket{u^{v}_{\mathbf{k}}}\rightarrow e^{i\varphi^{v}(\mathbf{k})}\ket{u^{v}_{\mathbf{k}}}\\
\label{eq:gauge_F}
&F(\mathbf{p},\mathbf{Q})\rightarrow e^{i(\varphi^{v}(\mathbf{p}-\mathbf{Q}/2)-\varphi^{c}(\mathbf{p}+\mathbf{Q}/2))}e^{i\theta(\mathbf{Q})}F(\mathbf{p},\mathbf{Q}).
\end{align}
Under this gauge transformation we have that $\mathbf{P}^{exc}_{p}(\mathbf{Q})=i F^{*}(\mathbf{p},\mathbf{Q})\mathbf{\nabla}_{\mathbf{p}}F(\mathbf{p},\mathbf{Q})+|F(\mathbf{p},\mathbf{Q})|^2 (\mathcal{A}^{c}(\mathbf{p},\mathbf{Q})-\mathcal{A}^{v}(\mathbf{p},\mathbf{Q}))$ transforms as:
{\allowdisplaybreaks
\begin{align}
\mathbf{P}_{\mathbf{p}}&\longrightarrow i e^{-i(\varphi^{v}(\mathbf{p}-\mathbf{Q}/2)-\varphi^{c}(\mathbf{p}+\mathbf{Q}/2))}e^{-i\theta(\mathbf{Q})}F^{*}(\mathbf{p},\mathbf{Q})\nonumber\\
&\Big((i\partial_{\mathbf{p}}\varphi^{v}(\mathbf{p}-\mathbf{Q}/2)-i\partial_{\mathbf{p}}\varphi^{c}(\mathbf{p}+\mathbf{Q}/2))\nonumber\\
&e^{i(\varphi^{v}(\mathbf{p}-\mathbf{Q}/2)-\varphi^{c}(\mathbf{p}+\mathbf{Q}/2))}e^{i\theta(\mathbf{Q})}F(\mathbf{p},\mathbf{Q})+\nonumber\\
&+e^{i(\varphi^{v}(\mathbf{p}-\mathbf{Q}/2)-\varphi^{c}(\mathbf{p}+\mathbf{Q}/2))}e^{i\theta(\mathbf{Q})}\mathbf{\nabla}_{\mathbf{p}}F(\mathbf{p},\mathbf{Q})\Big)\nonumber\\
&+|F(\mathbf{p},\mathbf{Q})|^2(-\partial_{\mathbf{k}^c}\varphi^{c}(\mathbf{k}^c)+\mathcal{A}^{c}(\mathbf{p},\mathbf{Q}))\nonumber\\
&+|F(\mathbf{p},\mathbf{Q})|^2(\partial_{\mathbf{k}^v}\varphi^{v}(\mathbf{k}^v)-\mathcal{A}^{v}(\mathbf{p},\mathbf{Q}))\nonumber\\
&= F^{*}(\mathbf{p},\mathbf{Q})\mathbf{\nabla}_{\mathbf{p}}F(\mathbf{p},\mathbf{Q}) + |F(\mathbf{p},\mathbf{Q})|^2(\mathcal{A}^{c}(\mathbf{p},\mathbf{Q})
\nonumber\\&\quad
-\mathcal{A}^{v}(\mathbf{p},\mathbf{Q}))-|F(\mathbf{p},\mathbf{Q})|^2(\partial_{\mathbf{k}^v}\varphi^{v}(\mathbf{k}^v)-\partial_{\mathbf{k}^c}\varphi^{c}(\mathbf{k}^c))
\nonumber\\&\quad
+|F(\mathbf{p},\mathbf{Q})|^2(\partial_{\mathbf{k}^v}\varphi^{v}(\mathbf{k}^v)-\partial_{\mathbf{k}^c}\varphi^{c}(\mathbf{k}^c))
\nonumber\\
&= F^{*}(\mathbf{p},\mathbf{Q})\mathbf{\nabla}_{\mathbf{p}}F(\mathbf{p},\mathbf{Q}) + |F(\mathbf{p},\mathbf{Q})|^2(\mathcal{A}^{c}(\mathbf{p},\mathbf{Q})
\nonumber\\&\quad
-\mathcal{A}^{v}(\mathbf{p},\mathbf{Q}))
\nonumber\\
&= \mathbf{P}_{\mathbf{p}}.
\end{align}}
This calculation demonstrates that the quantity $\mathbf{P}_{\mathbf{p}}$ is indeed gauge invariant within the considered gauge for the exciton problem. A similiar calculation shows that under the gauge transformation in Eqs.~\eqref{eq:gauge_uc},~\eqref{eq:gauge_uv},~\eqref{eq:gauge_F}, the Berry curvature ~\eqref{eq:Berry_Curvature_Exciton} is also gauge invariant.

\section*{Appendix C: Derivation of expression for \texorpdfstring{$\mathbf{P}_{p}$}{Pp} of the main text}

Here, we derive the expression for $\mathbf{P}_{p}$ in Eq.~\eqref{eq:P_p}. We start by recalling that the wavepacket of the exciton is given by:
\begin{equation}
  \Psi^{exc}_{\omega} = \sum_{\mathbf{Q}} \omega(\mathbf{Q}) \psi^{exc}_{\mathbf{Q}},
\end{equation}
with $\psi^{exc}{\mathbf{Q}} = e^{i\mathbf{Q}\cdot \mathbf{R}}\sum_{\mathbf{p}} F(\mathbf{p},\mathbf{Q})e^{i\mathbf{p}\cdot \mathbf{r}}u^{c}(\mathbf{p}+\mathbf{Q}/2)(u^{v}(\mathbf{p}-\mathbf{Q}/2))^{*}$ representing the exciton wavefunction. We substitute these expressions into the expectation value of the relative position operator $\hat{\mathbf{r}}$ and we will take the continuum limit of the center of mass momentum $\mathbf{Q}$, i.e. $\sum_{\mathbf{Q}}\rightarrow \frac{1}{(2\pi)^2}\int \mathrm{d}\mathbf{Q}$:
\begin{align}
\mathbf{r}_{\omega} &= \bra{\Psi^{exc}}\hat{\mathbf{r}}\ket{\Psi^{exc}}
= \frac{1}{(2\pi)^4}\!\int\! \mathrm{d}\mathbf{r}_c 
\!\int\! \mathrm{d}\mathbf{r}_v \sum_{\mathbf{p},\mathbf{p}'}
\!\int\! d\mathbf{Q}' \Bigl[
\nonumber\\&
\omega^{*}(\mathbf{Q}')e^{-i\mathbf{Q}'\cdot \mathbf{R}} F^{*}(\mathbf{p}',\mathbf{Q}') e^{-i\mathbf{p}'\cdot \mathbf{r}}
\nonumber\\&
(u^{c}_{\mathbf{p}+\mathbf{Q}/2}(\mathbf{r}_c))^{*}u^{v}_{\mathbf{p}-\mathbf{Q}/2}(\mathbf{r}_v)\hat{\mathbf{r}}
\!\int\!  \mathrm{d}\mathbf{Q} \omega(\mathbf{Q}) e^{i\mathbf{Q}\cdot \mathbf{R}} e^{i\mathbf{p}\cdot\mathbf{r}} 
\nonumber\\&
F(\mathbf{p},\mathbf{Q})
u^{c}_{\mathbf{p}+\mathbf{Q}/2}(\mathbf{r}_c)(u^{v}_{\mathbf{p}-\mathbf{Q}/2}(\mathbf{r}_v))^*\Bigr].
\end{align}
Above we introduced the dependence in the position of the periodic part Bloch wavefunction of the conductance and valence band explicitly, as we will perform integrals over the position. It is also important to note that $\mathbf{R}$ and $\mathbf{r}$ depend on the position in the conduction and valence band $\mathbf{r}_c$ and $\mathbf{r}_v$ according to the coordinate transformation of Sec.~\ref{sec:Exciton_Wavefunction_beyond_the_Effective_Mass_Approximation}.

We will now use the fact that the action of the relative position operator $\hat{\mathbf{r}}$ can be substituted by a derivative over the relative momenta in the following way:
\begin{equation}
\hat{\mathbf{r}}e^{i\mathbf{p}\cdot\mathbf{r}} = -i\frac{\partial}{\partial\mathbf{p}}e^{i\mathbf{p}\cdot\mathbf{r}}.
\end{equation}
In addition to this, we will take the continuum limit of the relative momenta $\mathbf{p}$, i.e. $\sum_{\mathbf{p}}\rightarrow \frac{1}{(2\pi)^2}\int \mathrm{d}\mathbf{p}$, this will be useful for the calculation, in the end we will go back to a sum over the relative momenta. We are then left with:
{\allowdisplaybreaks
\begin{align}
\mathbf{r}_{\omega} 
&= 
-\frac{i}{(2\pi)^8}\!\int\!\! \mathrm{d}\mathbf{r}_c \!\int\!\! \mathrm{d}\mathbf{r}_v \!\int\!\! \mathrm{d}\mathbf{p}' \!\int\!\! \mathrm{d}\mathbf{Q}' 
\Bigl[\omega^{*}(\mathbf{Q}')e^{-i\mathbf{Q}'\cdot \mathbf{R}}
\nonumber\\
&F^{*}(\mathbf{p}',\mathbf{Q}')
e^{-i\mathbf{p}'\cdot \mathbf{r}} 
(u^{c}_{\mathbf{p}+\mathbf{Q}/2}(\mathbf{r}_c))^{*}u^{v}_{\mathbf{p}-\mathbf{Q}/2}(\mathbf{r}_v)\!\int\!\! \mathrm{d}\mathbf{Q} \!\int\!\! \mathrm{d}\mathbf{p} 
\nonumber\\
&\omega(\mathbf{Q})
e^{i\mathbf{Q}\cdot \mathbf{R}}\hat{\mathbf{r}}e^{i\mathbf{p}\cdot\mathbf{r}}F(\mathbf{p},\mathbf{Q})u^{c}_{\mathbf{p}+\mathbf{Q}/2}(\mathbf{r}_c)(u^{v}_{\mathbf{p}-\mathbf{Q}/2}(\mathbf{r}_v))^{*}\Bigr]
\nonumber\\
&=\frac{i}{(2\pi)^8}\!\int\!\! \mathrm{d}\mathbf{r}_c \!\int\!\! \mathrm{d}\mathbf{r}_v \!\int\!\! \mathrm{d}\mathbf{p}' \!\int\!\! \mathrm{d}\mathbf{Q}' 
\biggl[\omega^{*}(\mathbf{Q}')e^{-i\mathbf{Q}'\cdot \mathbf{R}} 
\nonumber\\
&F^{*}(\mathbf{p}',\mathbf{Q}')
e^{-i\mathbf{p}'\cdot \mathbf{r}} 
(u^{c}_{\mathbf{p}+\mathbf{Q}/2}(\mathbf{r}_c))^{*}u^{v}_{\mathbf{p}-\mathbf{Q}/2}(\mathbf{r}_v)\!\int\!\! \mathrm{d}\mathbf{Q} \!\int\!\! \mathrm{d}\mathbf{p}
\nonumber\\
&\omega(\mathbf{Q}) e^{i\mathbf{Q}\cdot \mathbf{R}} e^{i\mathbf{p}\cdot\mathbf{r}}
\Big(\frac{\partial F(\mathbf{p},\mathbf{Q})}{\partial\mathbf{p}}u^{c}_{\mathbf{p}+\mathbf{Q}/2}(\mathbf{r}_c)(u^{v}_{\mathbf{p}-\mathbf{Q}/2}(\mathbf{r}_v))^{*}
\nonumber\\&\qquad+F(\mathbf{p},\mathbf{Q})\frac{\partial u^{c}_{\mathbf{p}+\mathbf{Q}/2}(\mathbf{r}_c)}{\partial\mathbf{p}}(u^{v}_{\mathbf{p}-\mathbf{Q}/2}(\mathbf{r}_v)
)^{*}+
\nonumber\\&\qquad+F(\mathbf{p},\mathbf{Q})u^{c}_{\mathbf{p}+\mathbf{Q}/2}(\mathbf{r}_c)\frac{\partial (u^{v}_{\mathbf{p}-\mathbf{Q}/2}(\mathbf{r}_v))^{*}}{\partial \mathbf{p}}\Big)\biggr].
\end{align}}
In the last step we integrated by parts in the relative momenta variable $\mathbf{p}$. The boundary term of the integration by parts is zero, because we are assuming that the exciton is a bound state and when the relative momenta tends to $\pm \infty$ it must vanish.

We have three integrals to solve above, in order to do that we will need to make use of the fact that the wavefunctions of the conduction, $\psi_{\mathbf{k}^c}=e^{i\mathbf{k}^c\cdot \mathbf{r}_c}u_{\mathbf{k}^c}(\mathbf{r}^c)$ and valence band, $\psi_{\mathbf{k}^v}=e^{i\mathbf{k}^v\cdot \mathbf{r}_v}u_{\mathbf{k}^v}(\mathbf{k}_v)$, are orthonormal, i.e.:
\begin{align}
    \braket{\psi_{\mathbf{k}'^c}|\psi_{\mathbf{k}^c}} &= \int \mathrm{d}\mathbf{r}_c e^{-i\mathbf{k}'^c.\mathbf{r}_c}u^{*}_{\mathbf{k}'^c}(\mathbf{k}_c)e^{i\mathbf{k}^c.\mathbf{r}_c}u_{\mathbf{k}_c}(\mathbf{r}_c)\nonumber\\
    \label{eq:Appendix_orthonomalization_u_c}
    &= \delta(\mathbf{k}^c-\mathbf{k}'^c),
\end{align}
\begin{align}
    \braket{\psi_{\mathbf{k}'^v}|\psi_{\mathbf{k}^v}} &= \int \mathrm{d}\mathbf{r}_v e^{-i\mathbf{k}'^v.\mathbf{r}_v}u^{*}_{\mathbf{k}'^v}(\mathbf{r}_v)e^{i\mathbf{k}^v.\mathbf{r}_v}u_{\mathbf{k}^v}(\mathbf{r}_v)\nonumber\\
    \label{eq:Appendix_orthonormalization_u_v}
    &= \delta(\mathbf{k}^v-\mathbf{k}'^v).
\end{align}
In addition to this, we will also use the fact that $\omega(\mathbf{Q})$ is a Gaussian envelope centered around $\mathbf{Q}=\mathbf{Q}_{\omega}$, which in turn means that integrals of smooth functions of $\mathbf{Q}$ and $\mathbf{p}$, here denoted $f(\mathbf{p},\mathbf{Q})$, will be:
\begin{align}
\sum_{\mathbf{Q}} \omega(\mathbf{Q}) f(\mathbf{p},\mathbf{Q}) = f(\mathbf{p},\mathbf{Q}_{\omega)}).
\label{eq:Appendix_integral_centered}
\end{align}
The result of the three integrals can be easily obtained using Eq.~\eqref{eq:Appendix_orthonomalization_u_c}, Eq.~\eqref{eq:Appendix_orthonormalization_u_v} and Eq.~\eqref{eq:Appendix_integral_centered}:
\begin{align}
\mathbf{r}_\omega(\mathbf{Q}_\omega) &= 
\sum_{\mathbf{p}}|F(\mathbf{p},\mathbf{Q_\omega})|^2 
\Bigl(\!
\mathcal{A}^{c}\bigl(\mathbf{p}+\tfrac{\mathbf{Q}_\omega}{2}\bigr) - 
\mathcal{A}^{v}\bigl(\mathbf{p}-\tfrac{\mathbf{Q}_\omega}{2}\bigr)\!\Bigr)\nonumber\\
&\qquad+i\sum_{\mathbf{p}} F^{*}(\mathbf{p},\mathbf{Q_\omega})\mathbf{\nabla}_{\mathbf{p}} F(\mathbf{p},\mathbf{Q_\omega}),
\end{align}
which is the result of Eq.~\eqref{eq:P_p} of the main text.


%

\end{document}